\documentclass[aps,amsmath,amssymb,preprintnumbers,a4paper,prd,twocolumn,nofootinbib]{revtex4-1}
\pdfoutput=1

\usepackage{amssymb,amsmath,latexsym,graphics, graphicx,epsfig,multirow,comment,hyperref,appendix, verbatim} 
\usepackage{pifont}
\usepackage{lipsum}
\usepackage{nicefrac}
\usepackage{slashed}
\usepackage{subfigure}
\usepackage{graphicx}
\usepackage{amsfonts}
\usepackage{color}
\usepackage{cancel}
\usepackage[normalem]{ulem}

\def\hbar{\hspace{0pt}\raisebox{1pt}{$-$} \hspace{-7pt} h}

\def\5{\overline 5}

\definecolor{JJ}{RGB}{0,144,255}

\newcommand{\be}{\begin{equation}}
\newcommand{\ee}{\end{equation}}
\newcommand{\bea}{\begin{eqnarray}}
\newcommand{\eea}{\end{eqnarray}}
\newcommand{\beq}{\begin{eqnarray}}
\newcommand{\eeq}{\end{eqnarray}}

\newcommand{\ba}{\begin{eqnarray}}
\newcommand{\ea}{\end{eqnarray}}

\newcommand{\vc}{F_{\mathrm{\pi}}}
\newcommand{\vf}{v}
\newcommand{\vw}{v_{\mathrm{w}}}

\newcommand{\SUL}{\mathrm{SU}(2)_{\mathrm{L}}}

\begin{document}
\title{Testing a dynamical origin of Standard Model fermion masses
}

\author{Tommi Alanne}
 \affiliation{CP$^{3}$-Origins, University of Southern Denmark, Campusvej 55, DK-5230 Odense M, Denmark}
\author{Diogo Buarque Franzosi}
 \affiliation{
 II. Physikalisches Institut, Universit\"at G\"ottingen, Friedrich-Hund-Platz 1, 37077 G\"ottingen, Germany
}
\author{Mads T. Frandsen}
 \affiliation{CP$^{3}$-Origins, University of Southern Denmark, Campusvej 55, DK-5230 Odense M, Denmark}

\begin{abstract}
We discuss a test of the Standard Model fermion mass origin in models of dynamical electroweak symmetry breaking. 
The couplings of composite pseudoscalar resonances to top quarks  allow to distinguish high-scale Extended-Technicolor-type fermion mass generation from fermion partial compositeness and low-scale mass generation via an induced vacuum expectation value of a doublet coupled to the composite sector. These different possible origins of fermion masses are thus accessible via weak-scale physics searched for at the LHC.

\end{abstract}
\preprint{CP3-Origins-2016-030 DNRF90}



\maketitle

\section{Introduction}\label{sec:seesaw}

In this paper, we consider  models of dynamical electroweak symmetry breaking (EWSB). We show that the observation of isoscalar CP-odd composite resonances $\eta_X$ not only sheds light on the underlying composite dynamics but also 
on the mechanism of mass generation for the Standard-Model (SM) fermions. In particular, the would-be Goldstone boson (GB) related to the $U(1)_A$ anomaly that we refer to as the $\eta_1$ state is interesting in this respect --- In QCD the analogous $\eta_1$ and $\eta_8$ states mix into the physical $\eta,\eta'$ states.  

If the top-quark mass arises from a direct coupling of the top quarks to the condensate in the strong sector, mediated by new high-scale states as in Extended-Technicolor (ETC) models \cite{Eichten:1979ah,Dimopoulos:1979es, 'tHooft:1979bh}, this implies Higgs-sized couplings of the new $\eta_1$ state to top quarks.  This is in contrast to the case where the SM-fermion masses arise via the so-called fermion partial compositeness (PC) mechanism \cite{Kaplan:1991dc}, or from a dynamically induced vacuum expectation value (vev) of a weak doublet, i.e. bosonic Technicolor (bTC). We note that such a doublet may be either elementary~\cite{Samuel:1990dq,Dine:1990jd} or composite~\cite{Chivukula:1990bc}.

In PC the $\eta_1$ couplings to top quarks are suppressed if the compositeness scale is higher than the electroweak scale and by the smallness of the underlying coupling between $\eta_1$ and the composite top partner, as discussed in Sec.~\ref{sec:PC}. 
In bTC the coupling is small because only the components of the doublet acquiring a vev are coupled directly to the SM fermions at interaction-eigenstate level, 
and the $\eta_1$ couples to the SM fermions only via mass mixing with the components of the doublet. 

The partial widths into $\bar{t}t$ versus $\gamma\gamma$ and $gg$ of the $\eta_1$ resonances thus provide an experimentally accessible diagnostic of the origin of SM-fermion masses.  
A complementary diagnostic of the fermion-mass origin is the momentum dependence of the Yukawa couplings above the mass-generation scale~\cite{Christensen:2005hm}.

\section{Effective Lagrangian description of the dynamical EWSB sector}
\label{sec:model}
We begin the discussion with a new sector able to break the electroweak symmetry dynamically and generate the observed $W$ and $Z$ masses \cite{Weinberg:1975gm,Susskind:1978ms}.  Minimally such a sector contains a single weak doublet of left-handed technifermions $Q_{\mathrm{L}}=(U_{\mathrm{L}}, D_{\mathrm{L}})$ and corresponding right-handed weak-singlet fermions $U_{\mathrm{R}},D_{\mathrm{R}}$ transforming in some representation 
under a new confining gauge group $G$. 
If the number of weak-doublet fermions is even, 
then the absence of gauge anomalies fixes the charges in the minimal case to $ Q(U)=1/2$, $Q(D)=-1/2$.
If the number of new weak doublet fermions is odd, then gauge anomalies can be satisfied by adding more matter and choosing the hypercharges appropriately, e.g.~\cite{Dietrich:2005jn}. 

Well below the condensation scale, we can describe the new strong sector  
using an effective Lagrangian invariant under a global symmetry group containing $SU(2)_{\mathrm{L}}\times SU(2)_{\mathrm{R}}$. Restricting to this subgroup, we write the Lagrangian in terms of a non-linear representation of the three GBs $\pi^{\pm,0}$ and $\eta_1$
via
\begin{equation}
\label{eq:sigdef}
\Sigma= \exp( i \pi/\vc), \,\,\,\,\,  \pi = \left( \begin{array}{cc} \eta_1 +  \pi^0 & \sqrt{2} \pi^+ \\ \sqrt{2}\pi^- 
    & \eta_1-\pi^0
\end{array}\right) \, .
\end{equation}
The transformation of $\Sigma$ under the global symmetry group is $\Sigma\to g_L \Sigma g_R^\dagger$ where $g_{L,R}\in SU(2)_{L,R}$.
The leading-order term of the effective Lagrangian is 
\begin{equation}
\mathcal{L}_{\rm K}=
 \frac{\vc^2}{4}{\rm Tr} [D_\mu^\dagger \Sigma D^\mu \Sigma],
\label{eq:LK}
\end{equation}
where $D_{\mu}$ is the electroweak covariant derivative,
while the anomaly-induced mass of the would-be GB $\eta_1$ can be encoded at the effective-Lagrangian level via the operator  
\begin{equation}
\mathcal{L}_{m} = \frac{m_{\eta_1}^2}{32}\vc^2{\rm Tr} [\ln \Sigma - \ln \Sigma^\dagger ]^2  \ .
\end{equation}

The $\eta_1$ state is particularly interesting, because its presence in the spectrum is model independent and its mass is sensitive to the compositeness scale  and fermion content of the composite sector  via $m_{\eta_1}^2 
\simeq \frac{N_f}{d(R)} \Lambda^2$, where $d(R)$ is the dimension of the constituent fermion representation under the strongly interacting gauge group, $N_f$ the number of flavours, and $\Lambda$ is a scale related to the GB decay constant and the dynamics-dependent anomaly term~\cite{Witten:1979vv,Veneziano:1979ec}\footnote{In the  Veneziano limit for $\mathrm{SU}(N)$ gauge group, $m_{\eta_1}^2\simeq 480  \frac{N_f}{N^2}\vc^2 $~\cite{DiVecchia:1980xq}.}. At the same time  its couplings are sensitive to the fermion mass mechanism. 

Nevertheless, in models with non-minimal global symmetries,  other isosinglet pseudo-GBs may be lighter, and the couplings of these states still probe the origin of fermion masses. We collectively denote these $\eta_X$.

\section{Origin of SM-fermion masses}

To generate SM-fermion masses, the condensate must be communicated to the SM fermions. 
One way is to extend the model with four-fermion interactions coupling two SM fermions, $q$, and two
technifermions, $Q$, via dimension-six operators of the schematic form $\mathcal{O}_4 \sim \frac{1}{\Lambda^2}q_L\bar{q}_RQ_L\bar{Q}_R$. These four-fermion operators can e.g. arise from exchanges of heavy spin-one fields~\cite{Eichten:1979ah,Dimopoulos:1979es} (after Fierz rearrangements) or heavy scalar doublets~\cite{'tHooft:1979bh}. In the following, we will refer to this as the Extended-Technicolor (ETC) scenario, independent of 
the origin of four-fermion operators.  

Another possibility is fermion partial compositeness, where dimension-six operators of the form $\mathcal{O}_4 \sim \frac{1}{\Lambda^2}q Q Q Q $ induce mixing of the SM fermions and composite baryons $B\sim QQQ$.

In these scenarios, such as technicolor or composite Higgs, a CP-even composite resonance could play the role of the observed Higgs boson. 

Another possibility is to extend the model with a light scalar doublet $\Phi$, analogous to the SM-Higgs doublet, coupled to the SM fermions and the new strongly-interacting fermions via Yukawa interactions. The doublet $\Phi$ may then obtain an induced vev from condensation and generate the SM-fermion masses~\cite{Samuel:1990dq,Dine:1990jd}. The neutral CP-even component of $\Phi$ may be interpreted as the 125-GeV Higgs boson.  The scalar doublet $\Phi$ may be fundamental and e.g. embedded in a supersymmetric theory at a higher scale~\cite{Samuel:1990dq,Dine:1990jd}, or composite and related to some additional composite dynamics above the weak scale~\cite{Chivukula:1990bc}.
We will refer to this scenario as bosonic technicolor (bTC) as in~\cite{Samuel:1990dq,Dine:1990jd}.

\subsection{Four-fermion operators (ETC)}
Upon integrating out the heavy ETC fields, the   $\SUL\times \mathrm{U}(1)_Y$-invariant effective operators responsible for generating the SM-fermion masses can be written as~\cite{DiVecchia:1980xq,Jia:2012kd,Molinaro:2016oix}
    \begin{equation}
	\label{eq:fermionmass}
	\begin{split}
	    \mathcal{L}_{\mathrm{ETC}}=&-Y_1 \vc f_1(h/\vc)\left(\bar{q}_{\mathrm{L}}\Sigma q_{\mathrm{R}}+\mathrm{h.c.}
		\right)\\
	    &-Y_2 \vc f_2(h/\vc)\left(\bar{q}_{\mathrm{L}}\Sigma \tau^3 q_{\mathrm{R}}+\mathrm{h.c.}
		\right) + \dots ,
	\end{split}
    \end{equation}
    where we restrict to the third-generation SM-quark doublet 
    $q_{\mathrm{L},\mathrm{R}}=(t,b)_{\mathrm{L},\mathrm{R}}$ and have absorbed the ETC scale in the Yukawa couplings, $Y_{1,2}$, and the dots represent additional interactions generated at low energies by the ETC interactions. The functions $f_{1,2}$ describe the excitations
    of the condensate,
        \begin{equation}
	\label{eq:}
	f_{1,2}=1+\kappa_{1,2}\frac{h}{\vc}+\dots
    \end{equation}
The lightest excitation, $h$, of the condensate may be interpreted as the 
125-GeV Higgs provided the unknown coefficients $\kappa_{1,2}$ reproduce the observed Higgs couplings within uncertainties. The $Y_{1,2}$ couplings are fixed by requiring the correct $t$ and $b$ masses: $m_t=(Y_1+Y_2)\vc$, and $m_b=(Y_1-Y_2)\vc$.
If the top mass is generated this way from ETC, then this also fixes the coupling between the quarks and $\eta_X$, see e.g.~\cite{DiVecchia:1980xq,Jia:2012kd,Molinaro:2016oix}.

An example of a composite sector featuring only a single doublet of 
technifermions is the $SU(3)_S$ 
Next-to-Minimal Walking Technicolor (NMWT) model~\cite{Sannino:2004qp,Dietrich:2005jn,Belyaev:2008yj} with $\eta_{\rm NMWT}  \sim \frac{1}{\sqrt{2}}(\bar{U}U+\bar{D}D)$. Then as emphasized in~\cite{Molinaro:2016oix}, we have
    \begin{equation}
	\label{eq:etaNMWT}
	\mathcal{L}_{\eta\bar{q}q}=-i\frac{m_t}{\vc}\eta_{\rm NMWT} \,\bar{t}\gamma^5t-i\frac{m_b}{\vc}\eta_{\rm NMWT} \,\bar{b}\gamma^5b , 
    \end{equation}
i.e. the top coupling is identical to that of the SM Higgs. 

An example of a model with a larger global symmetry group, and including colored constituents, is the so-called one-family model~\cite{Farhi:1980xs} for which ETC~\cite{Dimopoulos:1979es} and fermion partial compositeness~\cite{Kaplan:1991dc} extensions have both been explicitly constructed. 
In the one-family model, the diagonal $\eta_1$-type state (in analogy with QCD) is a mixture of colored and uncolored states and comes with an isosinglet partner, $\eta_{63}$, the analogue of the $\eta_8$ in the pseudoscalar nonet of QCD.

The anomaly coefficients $\mathcal{A}_{V_1 V_2}$, arising from the strongly-interacting constituents, that determine the dijet and diphoton decays of $\eta_X$ are defined as~\cite{Chivukula:2011ue}
\begin{align}
\mathcal{A}_{V_1V_2} = \mathrm{Tr}[T^a (T_1 T_2 + T_2 T_1)_{\mathrm{L}} + 
\mathrm{L}\leftrightarrow \mathrm{R}],
\end{align}
where $T^a$ is the generator of the axial-vector current associated with the state in question, and L,R denote the left- and right-handed underlying fermion components of the state. 
The anomaly coefficients and the reduced Yukawa couplings $c_X^{\rm ETC}\equiv y^{\mathrm{ETC}}_X \, \vw/ m_t $, induced by ETC,  for the one-family model 
are given by 

\bigskip
\begin{tabular}{ l c c c }
 state & $\quad A_{gg}$ & $\quad A_{\gamma\gamma}$  & $\quad c_X^{\mathrm{ETC}}$  \\
 \hline\hline
 $\eta_1\sim \frac{1}{4} (\bar{Q}i\gamma_5Q + \bar{L}i\gamma_5L)$ & $1$  & $\frac{8}{3}$& $1$ \\
 $\eta_{63}\sim   \frac{1}{4\sqrt{3}} (\bar{Q}i\gamma_5Q - 3 \bar{L}i\gamma_5L)$ & $\frac{1}{\sqrt{3}}$ & $\frac{-4}{3\sqrt{3}}$ & $\frac{1}{\sqrt{3}}$\\
\end{tabular}
\bigskip
In a generic model, due to the $\eta_X$ wave functions, the top-Yukawa couplings involve $\mathcal{O}(1)$ normalization factors as discussed below.

Returning to the low energy description of ETC induced mass terms in Eq.~\eqref{eq:fermionmass}, the Yukawa couplings $Y_{1,2}$ can e.g. be generated via integrating out a heavy scalar doublet~\cite{'tHooft:1979bh} with 
interactions
\begin{equation}
	    \label{eq:Yuk}
	    \begin{split}
		\mathcal{L}_{\mathrm{Yuk}}=&-y_t \bar{q}_{\mathrm{L}}\tilde{\Phi}t_{\mathrm{R}}
		    -y_b \bar{q}_{\mathrm{L}}\Phi b_{\mathrm{R}}\\
		&-y_U \bar{Q}_{\mathrm{L}}\tilde{\Phi} U_{\mathrm{R}}
		    -y_D \bar{Q}_{\mathrm{L}}\Phi D_{\mathrm{R}}\ +\ \mathrm{h.c.}
	    \end{split}
	\end{equation}
If we define the average coupling,  $Y \equiv \frac{1}{2} (y_U + y_D)$, and the relative difference, $\delta\equiv  \frac{y_U - y_D}{y_U + y_D}$, we can write 
$Y_1 \sim (y_t-y_b) Y \vc^2/m_{\Phi}^2$, and $Y_2 \sim (y_t+y_b) Y \vc^2/m_{\Phi}^2$. 
The $T$ parameter, measuring the amount of isospin breaking in the technicolor (TC) sector, 
is proportional to $\delta^2 Y^4$~\cite{Carone:1992rh}.
Therefore in the limit $y_U=y_D$, fermion masses can be generated without contributions to the $T$ parameter beyond those of the SM. 

Alternatively the interactions in Eq.~\eqref{eq:fermionmass} arise after Fierz-transforming vector-current four-fermion operators~\cite{Eichten:1979ah,Dimopoulos:1979es}. The discussion of the $\eta_X$ couplings is not affected by the difference in origin. 
    
\subsection{Induced vev (bTC)}

    Another possibility is that the scalar doublet in Eq.~\eqref{eq:Yuk} cannot be integrated out and obtains a vev, $\vf$. 
	To obtain the correct electroweak gauge boson masses, we require
	\begin{equation}
	    \label{eq:vw}
	    \vw^2 = N_D F_\pi^2 + \vf^2,
	\end{equation}
	with $\vw=246$~GeV, and $N_D$ is the number of families of condensing fermions.	The current measurements of the Higgs couplings constrain 
	$\sqrt{N_D}\vc\lesssim 0.4\, \vw$ if the scalar excitation of $\Phi$ is interpreted as the Higgs. On the other hand, it is reasonable to assume $\vc\gtrsim 0.25\, \vw$ not to have a very light resonance spectrum in disagreement with direct search constraints~\cite{Fodor:2016pls,Appelquist:2016viq}.~\footnote{The composite and elementary doublets can, in fact, mix kinetically, and this can modify Eq.~\eqref{eq:vw}, see e.g.~\cite{Alanne:2013dra}. This effect can help lifting the compositeness scale in bTC scenarios and therefore help evading the direct search constraints. Alternatively, the compositeness scale can be raised by vacuum misalignment mechanism as in composite Higgs scenarios.}

	The Yukawa terms in Eq.~\eqref{eq:Yuk} induce the following operators relevant to fermion masses 
	\begin{equation}
	    \label{eq:fermionmassvev}
	    \mathcal{L}_{\mathrm{mix}}=4\pi c_1\vc^3\left(y_U \mathrm{Tr}[U_{\Phi} \Sigma_U]+y_D \mathrm{Tr}[U_{\Phi} 
	    \Sigma_D]\right) 
	    + \text{h.c}
	\end{equation}
	where $c_1\sim\mathcal{O}(1)$ by naive dimensional analysis~\cite{Manohar:1983md,Georgi:1986kr}, 
	$U_{\Phi}=(\Phi\ \tilde{\Phi})$ is the $\SUL\times \mathrm{SU}(2)_{\mathrm{R}}$ bidoublet matrix, 
	and we have defined the projections 
	\begin{equation}
	    \label{eq:}
	    \begin{split}
		\Sigma_U\sim\left(\begin{array}{cc}U_LU_R^{\dagger}&0\\D_LU_R^{\dagger}&0\end{array}\right),\
		\Sigma_D\sim\left(\begin{array}{cc}0&U_LD_R^{\dagger}\\0&D_LD_R^{\dagger}\end{array}\right).
	    \end{split}
	\end{equation}
	The above terms generate a mass mixing between the pseudoscalar triplet $\pi_f$ contained in $\Phi$, the composite triplet of `pions'  $\pi_c$ in Eq.~(\ref{eq:sigdef}) and $\eta_X$.
	The (mostly-)$\eta_X$ mass eigenstate couples to the top quark only via this mass mixing with $\pi_f^0$ which is 
	absent in the $y_U=y_D$ limit of unbroken isospin in the composite sector.
	To lowest order in the isospin-breaking parameter $\delta$,
	we find 
	\begin{equation}
	    \label{eq:}
	    \begin{split}
		c_X &\approx 8 \sqrt{2}\pi c_1 Y \delta \frac{\vc^2}{m_{\eta_X}^2} \frac{\vw}{v} +\mathcal{O}(\delta^2) , 
	    \end{split}
	\end{equation}
	where $c_X \equiv y_X \, \vw/ m_t $ is the $\eta_X$ Yukawa coupling in units of the SM Higgs Yukawa coupling. 
	Similarly as in the case with ETC, in extended models the $\eta_X$ wave functions involve $\mathcal{O}(1)$ normalization factors that
	can further suppress the top coupling.
	The amount of isospin breaking in the TC sector is highly constrained by the $T$ parameter, which
	in the limit where the elementary scalar is much lighter than the compositeness scale, $m_{\Phi}\ll\Lambda$, 
	is proportional to $\delta^2Y^2$~\cite{Carone:1992rh}.

Fixing $m_{\Phi}=125$~GeV, and $c_1=1$, we find that requiring $T$ parameter within $1\sigma$ confidence 
level from the experimental value constraints $\delta\cdot Y\lesssim 0.2$ for $\vc>0.25 \vw$. 

We also note that the physical triplet of CP-odd states couples to top quarks only through its small elementary component. Therefore also this top
coupling is very suppressed, and these additional pseudoscalar states do not spoil our ability to distinguish this scenario from ETC induced fermion masses. However, their mass and properties
are further constrained  by data and vacuum stability~\cite{Carone:2012cd}.

\vspace{0.5cm}

\subsection{Top partners (PC)}
\label{sec:PC}

A third possibility to generate masses for the SM fermions, is via mixing with composite baryons~\cite{Kaplan:1991dc}. 
Then the low-energy effective theory with the composite baryon $B\sim QQQ$ contains the following operators:
\begin{equation}
    \label{eq:toppartner}
    \mathcal{L}_{\mathrm{eff}}=\epsilon\left(q B+\bar{q}\bar{B}\right)-m_BB\bar{B}+\,\mathrm{h.c.}
\end{equation}
and diagonalization gives rise to the quark mass $q$.

The composite interaction $C_X \frac{m_B}{\vc}\eta_XB\bar{B}$ generates the coupling between the $\eta_X$ state and the SM top quark after diagonalization. The result is given by
\begin{equation}
    \label{eq:}
    c_X\approx\frac{C_X \vw}{\vc},
\end{equation}
where $\vc$ is the compositeness scale and $C_X$ includes the form factor between the $\eta_X$ and the technibaryon and the normalisation factor of the state. In composite Higgs models, where fermion partial compositeness has been extensively employed, the electroweak scale arises due to vacuum misalignment such that $\vw=\vc\sin\theta$, and  $\theta\ll 1$ to avoid electroweak precision test constraints. The above top coupling is therefore significantly smaller than in the ETC case.

Moreover, in QCD the coupling between proton and the $\eta_1$ related to the axial anomaly is small~\cite{Veneziano:1989ei,Moskal:1998pc} implying that $C_1\ll 1$.   
Therefore, for top-quark partial compositeness in composite Higgs models, the coupling to axial $\eta_1$ is expected to be suppressed compared to the ETC case by both the higher compositeness scale and the small $\eta_1 B\bar{B}$ coupling.

\section{Experimental test of the fermion-mass mechanism}

We have seen that the top coupling of the isosinglet CP-odd 
resonance, $\eta_X$, in the new gauge sector is essentially binary with respect to the fermion-mass mechanisms considered: In the case of high-scale ETC-type mass generation, the top coupling 
is SM-Higgs sized. Instead, when a doublet acquiring a vev or mixing with top partners is responsible for fermion masses, the top coupling of $\eta_X$ is expected to be much smaller---ultimately limited by the contribution to the $T$ parameter or the compositeness scale.  Thus, this state presents a sensitive probe of the fermion-mass mechanism in models of dynamical EWSB.

Following the notation in~\cite{Djouadi:2005gj}, the partial widths for subsequent decays into top quarks,  photons and gluons are then given by 
\bea
\Gamma_{\eta_X \to \bar{t}t} &=&   \frac{3}{8\pi} 
    \frac{m_t^2}{\vw^2}c_X^2\,  m_{\eta_X}\left(1-m_t^2/m_{\eta_X}^2\right)^{1/2}  \nonumber
\\
\Gamma_{\eta_X\to \gamma\gamma} &\simeq&
\frac{\alpha^2m_{\eta_X}^3
}{256\pi^3\vw^2} \left| \frac{4}{3} c_{X} A^{A}_{1/2}(\tau_t) + \frac{N_{\mathrm{TC}}
\vw}{F_\pi}\mathcal{A}_{\gamma\gamma}^{\eta} \right|^2 \nonumber
\\
\Gamma_{\eta_X \to gg} &\simeq&  \frac{\alpha_s^2m_{\eta_X}^3
}{128 \pi^3\vw^2} \left|  c_X A^{A}_{1/2}(\tau_t)
+ \frac{2 N_{\mathrm{TC}}
\vw}{F_\pi}\mathcal{A}_{gg}^{\eta} \right|^2
\label{Eq:partials}
\eea
where we neglect the small contribution of light fermions and we fix the strong-interaction scale by Eq.~\eqref{eq:vw}.
 The loop function $A_{1/2}^{A}(\tau_f)$ is given in~\cite{Djouadi:2005gj} (Eq. (2.26)). 

Relative to a SM Higgs at a given mass $m_{h_{\rm SM}}$, the production cross section of $\eta_X$ is~\cite{Chivukula:2011ue}
\begin{align}
\frac{\sigma_{gg\to \eta_X}}{\sigma_{gg\to h_{\rm SM}}} = \frac{\Gamma(\eta_X\rightarrow gg)}{\Gamma(h_{\mathrm{SM}}\rightarrow gg)} \equiv R_{gg}.
\label{Eq:cs-scaling}
\end{align}
If some of the $\eta_X$ constituents carry color, and they dominate the gluon-fusion production we obtain
\begin{align}
R_{gg}\simeq 4 N_{\mathrm{TC}}^2 (A_{gg}^{\eta})^2Ê\frac{\vw^2}{\vc^2}  \left|A^{\mathcal{H}}_{1/2}(\tau_t) \right|^{-2},
\label{Eq:prodapprox}
\end{align}
where $A^{\mathcal{H}}_{1/2}(\tau_f)$ is given in~\cite{Djouadi:2005gj} (Eq. (2.25)).

The $\eta_X$-like state may also be produced in association with other states, e.g. via the decays of spin-1 isosinglet or isotriplet resonances like the analogues of the QCD $\omega$ and $\rho$ mesons. This requires these states to be heavy enough that the decays $\omega\to \eta_X \gamma$ or $\rho \to \eta_X \gamma$ proceed on shell, see e.g \cite{Carone:1993vg}. In this study we focus on the gluon-fusion production.

\begin{figure}
    \begin{center}
	\includegraphics[width=0.4\textwidth]{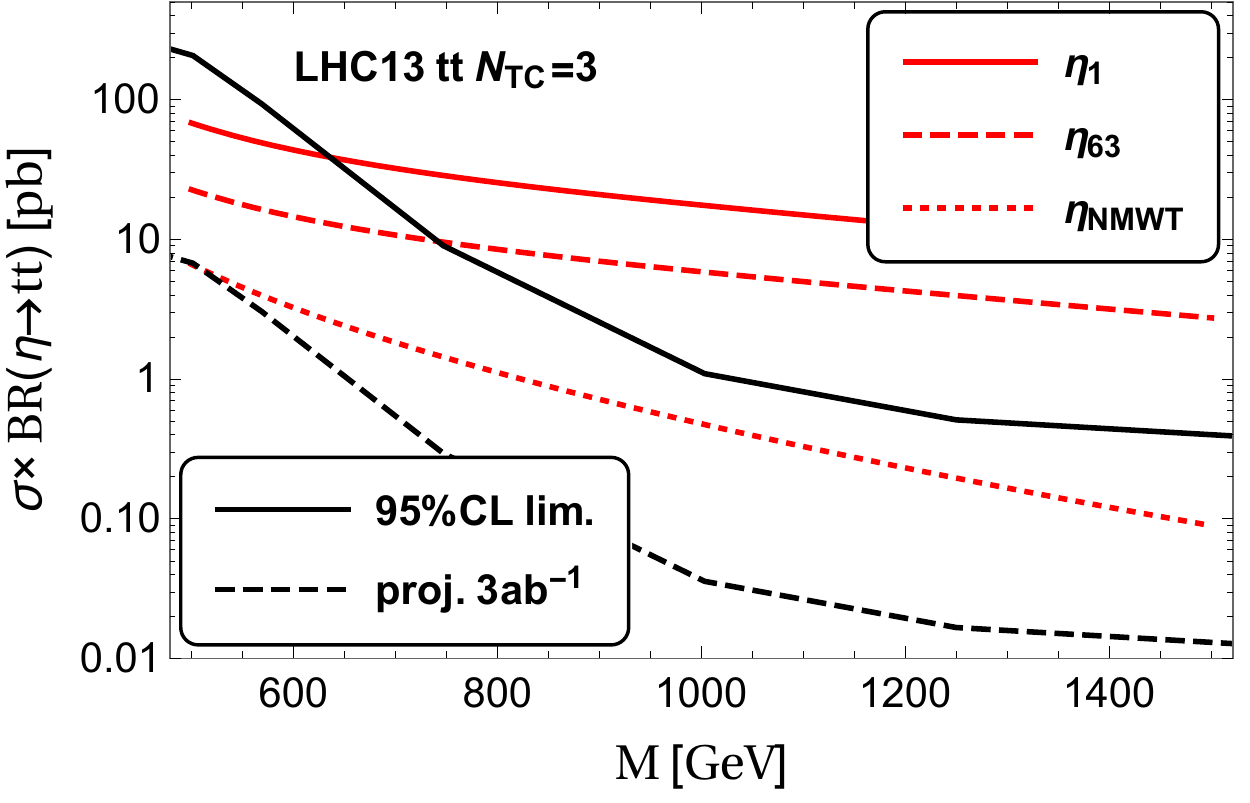}\\
	\vspace{0.2cm}
	\includegraphics[width=0.4\textwidth]{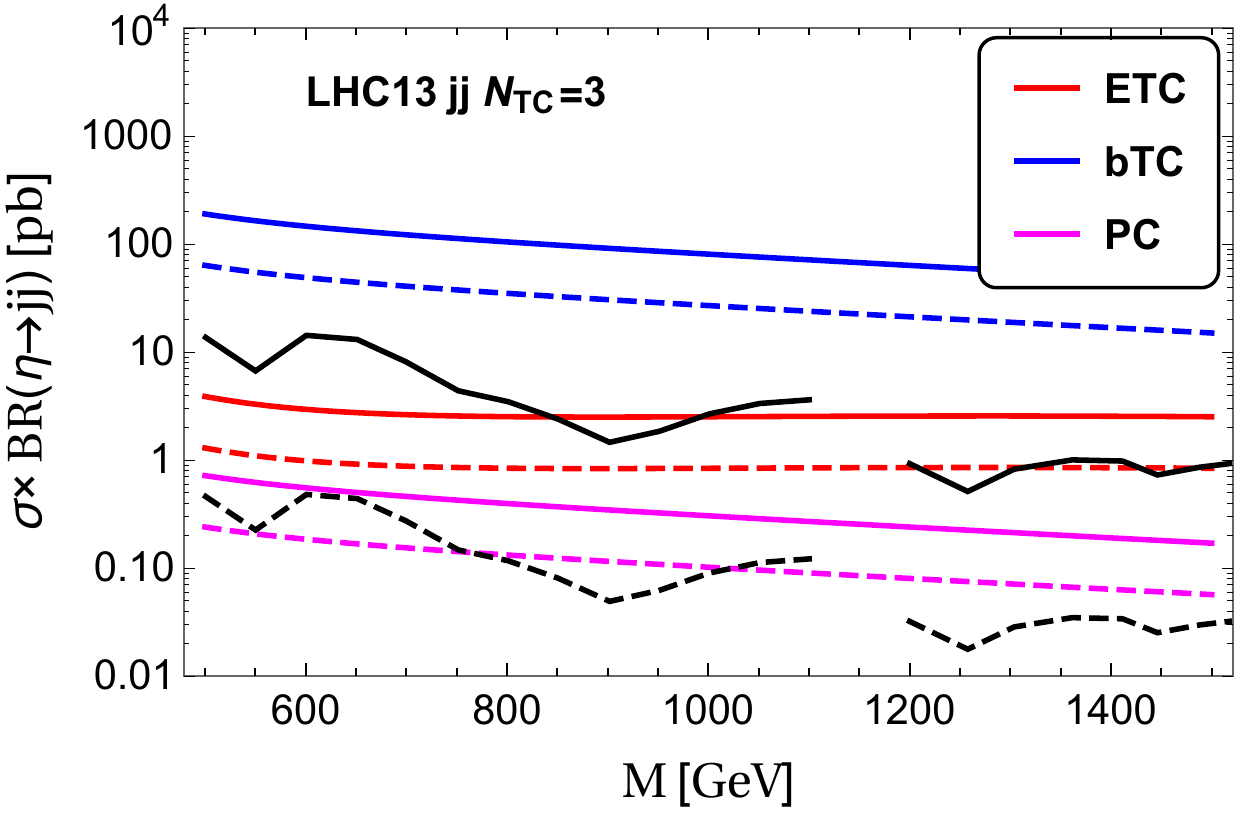}\\
	\vspace{0.2cm}
	\includegraphics[width=0.4\textwidth]{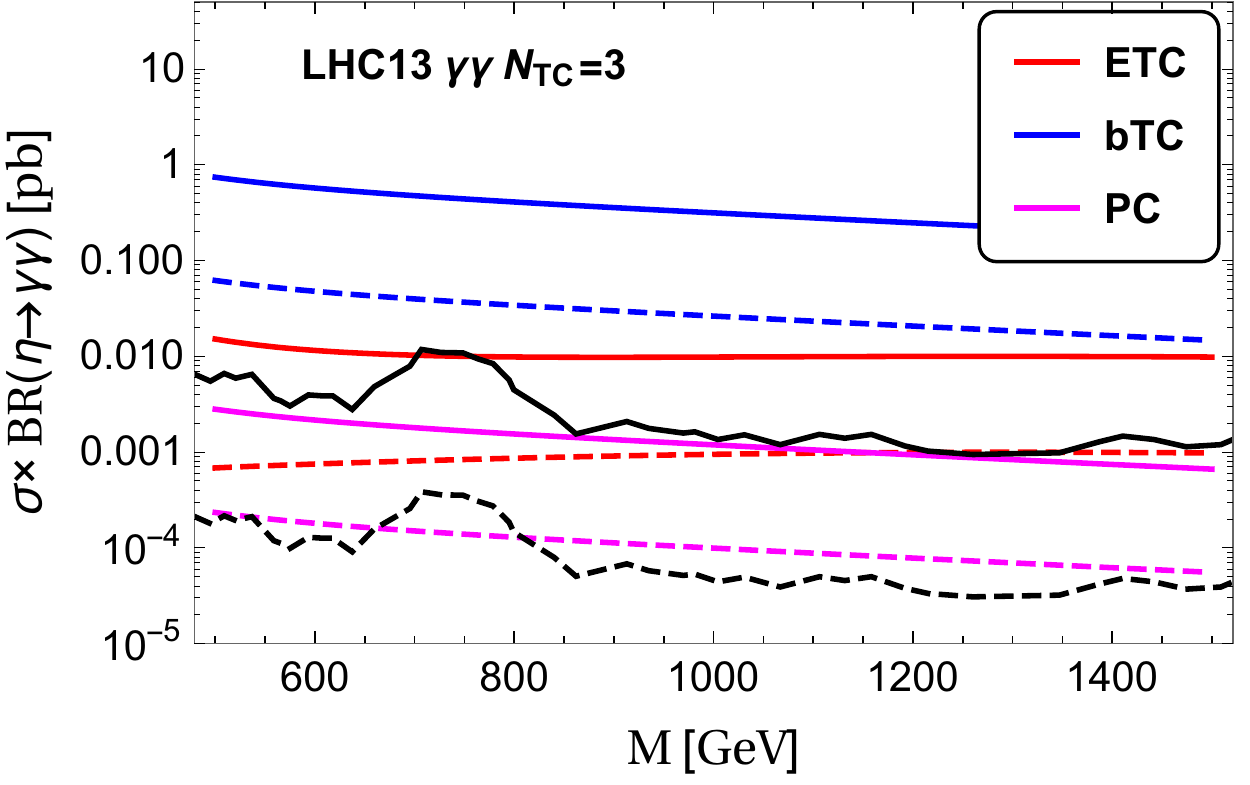}
    \end{center}
    \caption{Production cross section times branching rations of the $\eta_X$ states when ETC (red lines, $c_X^{\rm ETC}$), bTC (blue lines, $c_X=0$) and PC (magenta lines, $c_X=0$) is responsible for the top quark mass. Also shown are the LHC limits (solid black lines) and reach projections (black dashed lines):  {\bf Top:} $t\bar{t}$ channel.     {\bf Middle:} Dijet channel. {\bf Bottom:} Diphoton channel. 
   We use $N_{\mathrm{TC}}=3$ as a benchmark value and fix $v^2=\frac{3}{4} \vw^2$ for the one-family model ($F_\pi=\vw/4$),  while $v=0$ for the NMWT model (dotted line)  ($F_\pi=\vw)$. For PC, we use $F_{\pi}=1$ TeV as a benchmark value. We have omitted the NMWT case in the dijet and diphoton channels for clarity, since the cross section is tiny.    }
       \label{fig:xs}
\end{figure}
\bigskip

The diagnostic of ETC interactions, given discovery of an $\eta_X$ resonance is therefore a relatively broad resonance dominated by the top decay mode.

The $\bar{t}t$  cross sections of the $\eta_X$ states from the one-family model and the NMWT model are displayed in the top panel of Fig.~\ref{fig:xs} assuming the top mass arises from ETC-type interactions. Since in this case $\Gamma_{\eta_X\to tt}\simeq \Gamma_{\eta_X}$, the curves also represent the production cross sections of the $\eta_X$ states to a good approximation.  
We used the N$^3$LO prediction for the scalar production through gluon fusion~\cite{Anastasiou:2016hlm} and rescaled by $R_{gg}$.
The figure also shows the 95\% CL limit on the $\bar{t}t$ production cross section of a heavy resonance 
provided by the ATLAS analysis of highly boosted tops decaying semi-leptonically \cite{ATLAS-CONF-2016-014} with luminosity $L_0=3.2$fb$^{-1}$ (black solid curve). 
The limit indicates that the LHC is already able to rule out the ETC scenario in the $t\bar{t}$ channel in some composite scenarios, but our analysis does not include interference effects which are known to be important even for widths $\Gamma_{\eta_X}/m_{\eta_X} \gtrsim  2\%$ \cite{Hespel:2016qaf,Bernreuther:2015fts,Dicus:1994bm,Frederix:2007gi,Jung:2015gta} as here.

The middle and lower panels of Fig.~\ref{fig:xs} show production cross section times branching ratio into dijets and diphotons respectively,  
for the ETC (red), the bTC (blue), and the PC case (magenta). 

For the bTC case we take $v^2=(3/4)\vw^2$ to satisfy the observed Higgs decay rates.
For a four-doublet model such as the one-family model, this implies a very low compositeness scale $F_\pi=\vw/4$
due to the constraint in Eq.~(\ref{eq:vw}). For PC, the cross sections are suppressed by the compositeness scale, and we use $F_{\pi}=1$ TeV as a benchmark value to avoid constraints from electroweak precision tests in the composite-Higgs case. We note that the $\eta_1$ in this case might be too heavy for LHC reach, and a better probe for our diagnostic may be provided by a pseudo-GB, or  a more energetic future collider is necessary. 

In the middle panel, we also show the ATLAS 95\%\,CL exclusion limit on dijet resonances for low  
\cite{ATLAS-CONF-2016-030} and high 
\cite{ATLAS:2015nsi} resonance mass searches. Since the widths of our $\eta_X$ resonances satisfy $\Gamma_{\eta_X}/m_{\eta_X} \lesssim  5 \times 10^{-2}$, we used the exclusion limit relevant for a width equal to the detector mass resolution (referred to as the `Res.' analysis in \cite{ATLAS:2015nsi}). We also assumed (conservatively) an acceptance of $30\%$.
Similarly in the lower panel, we show the diphoton cross sections limits from ATLAS 
in \cite{Aaboud:2016tru} which assumes a narrow resonance with $\Gamma/m=2 \times 10^{-2}$. 
The integrated luminosities used in the dijet (high mass), dijet (low mass) and diphoton searches are respectively $L_0=3.4,\,3.6,\,3.2$fb$^{-1}$. 
Our projections for the cross-section limits from a very high luminosity (VHL) search with $L=3\,{\rm ab}^{-1}$ are obtained by a simple rescaling of the current limit by $\sqrt{L_0/L}$ and depicted as the dashed black lines.
It can be seen that in the one-family model set-up both ETC and bTC mechanisms are excluded and only PC may generate fermion masses and will be probed in the near future for scales $\vc$ up to 1 TeV. Other UV realizations should be studied.

Finally, for a minimal model without colored constituents, such as the $SU(3)_S$ NMWT example above, the gluon-fusion cross section is induced only via  the top quark.  In the ETC case, the production cross section shown with a dotted line in the top panel of Fig.~\ref{fig:xs} is then of the order of the SM Higgs with similar mass. This is below current sensitivity in the $\bar{t}t $ channel but the VHL LHC could be able to probe this state in the future. The associated production channels from heavy resonance decays mentioned above will also be relevant to probe this scenario.

\section{Conclusions}

There are multiple possible realizations of the SM fermion masses within the framework of dynamical EWSB. In particular, the scale of fermion-mass generation may be much higher than the weak scale as in ETC and fermion PC models. It is therefore very important to establish experimental probes able to test the fermion-mass generation in models with dynamical EWSB.

Here we have shown that a ubiquitous feature of weak-scale dynamical EWSB within reach of the LHC experiment, i.e. isoscalar CP-odd resonances, also provides experimental access to the origin of SM-fermion masses, notably the top-quark mass.  
We have shown that the value of the SM-top couplings for these resonances is very sensitive to the underlying fermion-mass-generation mechanism:
If high-scale interactions generate four-fermion operators that upon EWSB provide the top-quark mass, 
these resonances acquire an $\mathcal{O}(1)$ Yukawa coupling to the top quark like the SM Higgs~\cite{Jia:2012kd,Molinaro:2016oix}. Instead in models where the top mass arises from an induced vev of a scalar doublet, 
the top coupling is either $\ll 1$ or correlated with a sizable contribution to the $T$ parameter. In composite Higgs models with fermion partial compositeness it is suppressed by the high compositeness scale. The partial width of these resonances into $\bar{t}t$ relative to partial widths into $\gamma\gamma$ and $gg$ therefore offers a diagnostic of the underlying fermion-mass mechanism.

\acknowledgements
The CP$^3$-Origins centre is partially funded by the Danish National Research Foundation, grant number DNRF90.
Tommi Alanne and Mads T. Frandsen acknowledges partial funding from a Villum foundation grant.

\bibliography{PCH}

\end{document}